%% file: main.tex
\documentclass[letterpaper]{article} 
\usepackage{aaai2026}  
\usepackage{times}  
\usepackage{helvet}  
\usepackage{courier}  
\usepackage[hyphens]{url}  
\usepackage{graphicx} 
\usepackage{natbib}  
\usepackage{caption} 
\usepackage{algorithm}
\usepackage{algorithmic}
\usepackage{booktabs}
\usepackage{amsmath} 
\usepackage{color}
\usepackage{amssymb}
\usepackage{newfloat}
\usepackage{listings}
\DeclareCaptionStyle{ruled}{labelfont=normalfont,labelsep=colon,strut=off} 
\floatstyle{ruled}
\newfloat{listing}{tb}{lst}{}
\floatname{listing}{Listing}
\title{InstructDubber: Instruction-based Alignment for Zero-shot Movie Dubbing}
\author{
    Zhedong Zhang\textsuperscript{\rm 1,2}\footnote{This work is done during the intern in VIPL group, ICT, CAS.}, Liang Li\textsuperscript{\rm 2}\thanks{Corresponding author.}, Gaoxiang Cong\textsuperscript{\rm 2,3}, Chunshan Liu\textsuperscript{\rm 1}, Yuhan Gao\textsuperscript{\rm 1}, \\ 
    Xiaowan Wang\textsuperscript{\rm 4}, Tao Gu\textsuperscript{\rm 5}, Yuankai Qi\textsuperscript{\rm 5}
}
\affiliations{

\textsuperscript{\rm 1}Hangzhou Dianzi University, Hangzhou, China,\\
\textsuperscript{\rm 2}Institute of Computing Technology, Chinese Academy of Sciences, Beijing, China,\\
\textsuperscript{\rm 3}University of Chinese Academy of Science, Beijing, China,\\
\textsuperscript{\rm 4}Tsinghua University, Beijing, China,\\
\textsuperscript{\rm 5}Macquarie University, Sydney, Australia  \\
}

\usepackage{bibentry}

\begin{document}

\def\eg{\emph{e.g.}} 
\def\Eg{\emph{E.g.}}
\def\vs{\emph{v.s.}} 
\def\ie{\emph{i.e.}} 
\def\Ie{\emph{I.e.}}
\def\etc{\emph{etc.}} 
\def\wrt{\emph{w.r.t.}} 
\def\etal{\emph{et al.}}

\maketitle

\input{Sections/0-Abstract}

\begin{links}
    \link{Demo}{https://zzdoog.github.io/InstructDubber/}
\end{links}

\input{Sections/1-Intro}
\input{Sections/2-Related_Works}
\input{Sections/3-Method}

\input{Sections/4-Experiment}

\input{Sections/5-Conclusion}

\bibliography{aaai2026}


\end{document}

%% file: Sections/0-Abstract.tex
\begin{abstract}
Movie dubbing seeks to synthesize speech from a given script using a specific voice, while ensuring accurate lip synchronization and emotion-prosody alignment with the character’s visual performance. 
However, existing alignment approaches based on visual features face two key limitations: 
(1) they rely on complex, handcrafted visual preprocessing pipelines, including facial landmark detection and feature extraction; and (2) they generalize poorly to unseen visual domains, often resulting in degraded alignment and dubbing quality.
To address these issues, we propose InstructDubber, a novel instruction-based alignment dubbing method for both robust in-domain and zero-shot movie dubbing.
Specifically, we first feed the video, script, and corresponding prompts into a multimodal large language model to generate natural language dubbing instructions regarding the speaking rate and emotion state depicted in the video, which is robust to visual domain variations.
Second, we design an instructed duration distilling module to mine discriminative duration cues from speaking rate instructions to predict lip-aligned phoneme-level pronunciation duration.
Third, for emotion-prosody alignment, we devise an instructed emotion calibrating module, which fine-tunes an LLM-based instruction analyzer using ground truth dubbing emotion as supervision and predicts prosody based on the calibrated emotion analysis.
Finally, the predicted duration and prosody, together with the script, are fed into the audio decoder to generate video-aligned dubbing. 
Extensive experiments on three major benchmarks demonstrate that InstructDubber outperforms state‑of‑the‑art approaches across both in‑domain and zero‑shot scenarios.
\end{abstract}

%% file: Sections/1-Intro.tex
\section{Introduction}
Movie Dubbing, also known as Visual Voice Cloning (V2C)~\cite{V2C}, aims to transfer the given script into speech with a specific voice, while preserving temporal synchronization with the character’s lip movements and emotional alignment with their facial expressions in the video.
It has broad real-world applications in areas such as film production, digital media, and personalized speech AIGC.
However, the requirement of accurately and fine-grainedly aligning the speech with visual performance also presents a substantial challenge to movie dubbing task.

\begin{figure}[t]
    \centering
    \includegraphics[width=1\linewidth]{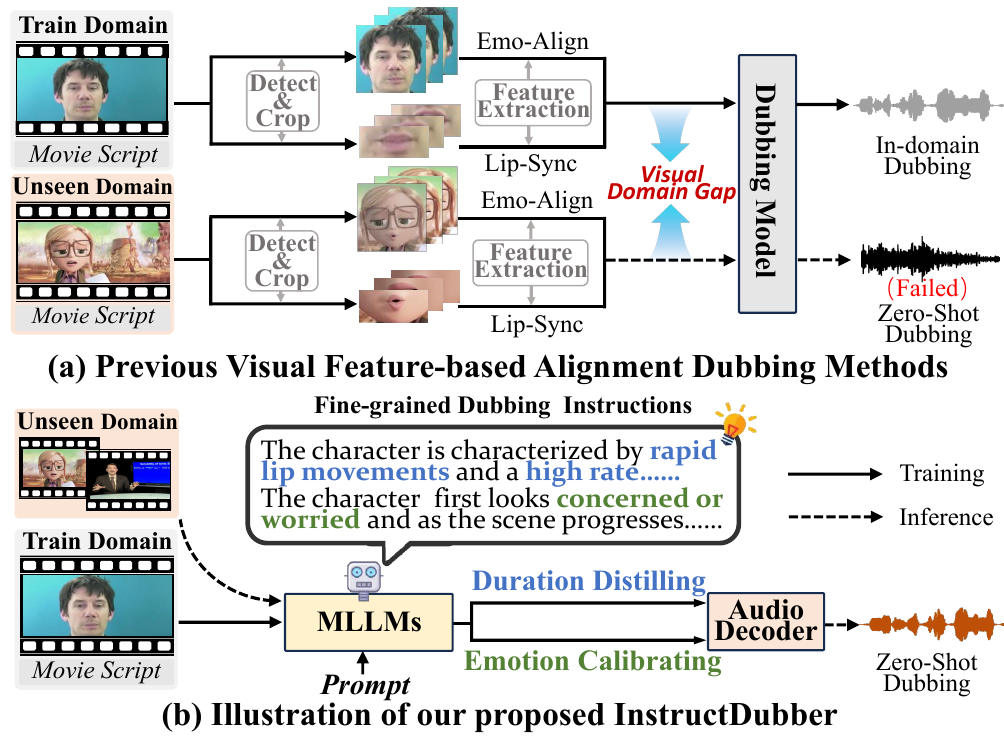}
    \caption{(a) Illustration of the previous dubbing methods with visual feature-based alignment, which rely on complex visual preprocessing and suffer from poor generalization to unseen visual domains.
             (b) Illustration of our proposed InstructDubber, an instruction-based alignment dubbing method that achieves robust zero-shot dubbing using natural language fine-grained dubbing instructions.
             }
    \label{Fig-intro}
\end{figure}

Existing alignment methods for movie dubbing broadly fall into two main categories.
The first category~\cite{nerualdubber, HPMDubbing,mcdubber} generates dubbing using the fusion representation of visual features from lip regions and the textual features from the script to achieve temporal alignment.
It improves the synchronization between the generated dubbing and lip movements but struggles to build clear speech from the lips of various visual scenes, thus often results in suboptimal speech quality.
The second category~\cite{styledubber, spk2dub} employs phoneme-based speech synthesis models~\cite{fs2, StyleTTS2} as the generation backbone, leveraging visual features of lip movements and facial expressions to predict video-aligned phoneme-level duration and prosodic attributes (\ie, pitch and energy).
By incorporating acoustic pre-training techniques~\cite{ProDubber}, the second category achieves improved video-dubbing alignment while maintaining high speech quality.

Despite the progress, the aforementioned approaches typically rely on complex and time-consuming handcrafted visual preprocessing steps like detecting and segmenting characters' facial and lip regions, followed by various feature extraction procedures, as illustrated in Figure~\ref{Fig-intro} (a).
Beyond the burdensome preprocessing pipeline, these visual-based alignment methods are sensitive to variations in visual domains, such as the domain gap between animated and live-action characters.
Consequently, the reliance on such preprocessed visual features makes these methods vulnerable to performance degradation when facing videos from unseen domains, severely compromising both alignment accuracy and dubbing quality.


Compared to visual features, natural language instructions serve a more intuitive and interpretable modality for dubbing alignment.
They provide fine-grained alignment guidance while offering better universality across diverse visual domains.
Meanwhile, the powerful multimodal understanding capabilities of the multimodal large language models (MLLMs) make it possible to generate visual-domain-robust fine-grained dubbing instructions directly from video input.
However, related prior methods only leverage coarse-grained instructions (\eg, character gender or age) as a supplement to visual features~\cite{deepdubber} or solely adopt the autoregressive framework for dubbing generation~\cite{voicecraftdub}, overlooking the capability of the instruction-based dubbing alignment and its zero-shot potential to generalize across diverse visual domains.

To this end, we propose InstructDubber, a novel instruction-based alignment dubbing method that effectively leverages fine-grained dubbing instructions to achieve robust in-domain and zero-shot movie dubbing (as shown in Figure~\ref{Fig-intro} (b)).
Specifically, we first feed both the video and script to a pre-trained MLLM to generate fine-grained, visual-domain-robust natural language dubbing instructions that capture characters' speaking rates and emotions using corresponding prompts.
Second, we propose an Instructed Duration Distilling module to mine duration cues from the speaking rate instruction. 
This is achieved by a set of learnable duration prototypes with slot-attention-based distillation.
The distilled duration cues are then used to predict the phoneme-level pronunciation duration together with prosodic text features of input script.
Third, for the emotion-prosody alignment, we propose an Instructed Emotion Calibrating module.
It fine-tunes a lightweight LLM to analyze the emotion instructions by leveraging emotion entities extracted from ground truth dubbing as supervision.
Based on the calibrated emotion entities extracted from emotion instruction by the fine-tuned analyzer, we predict the emotion-aligned prosody of each phoneme.
Finally, the script text features, combined with duration and prosody inferred from visual-domain-robust instructions, are provided to an audio decoder to synthesize temporally aligned, high-fidelity dubbing in both in-domain and zero-shot dubbing scenarios.



The main contributions are summarized as follows:
\begin{itemize}
\item We propose InstructDubber, a dubbing method with instruction-based alignment that effectively leverages natural language instructions to generate video-aligned dubbing in both in-domain and zero-shot scenarios.
\item We design an instructed duration distilling module to mine duration cues from fine-grained speaking rate instructions by slot-attention-based distillation to predict the lip-aligned phoneme-level pronunciation duration.
\item We devise an instructed emotion calibrating module that optimizes the analysis of fine-grained emotion instructions and models emotion-aligned dubbing prosody based on the calibrated emotion analysis.
\item Favorable performance on both in-domain and zero-shot dubbing scenarios across three major benchmarks demonstrates the effectiveness of our approach.
\end{itemize}

\begin{figure*}[t]
 \centering
  \resizebox{\linewidth}{!}{\includegraphics{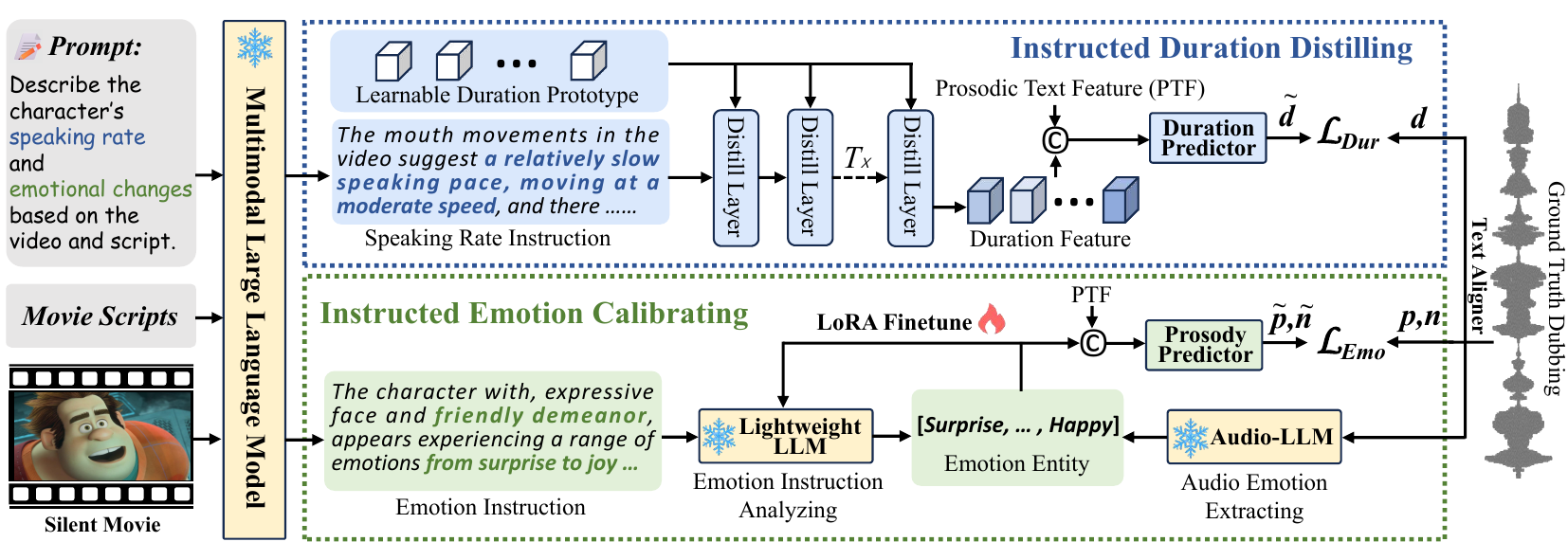}}
  \caption{
  The main architecture of the proposed InstructDubber.
  To predict the lip-synchronized phoneme-level duration, the Instructed Duration Distilling module (IDD) mines the duration cues from fine-grained speaking rate instructions.
  The Instructed Emotion Calibrating module (IEC) fine-tunes a lightweight LLM to analyze the emotion instructions using the emotion entities from ground truth dubbing as supervision, and predicts the dubbing prosody based on the calibrated emotion analysis.
  }
  \label{method}
\end{figure*}

%% file: Sections/2-Related_Works.tex
\section{Related Works}

\subsection{Speech Synthesis} 
With the rapid development of deep learning~\cite{DeT, zhao2025heterogeneous, yin2025progressive, chen2024sdpl,zhang2024inductive,tu2024smart, li2022long, zhang2025generating}, the FastSpeech series~\cite{fs2} first introduces a phoneme-level duration-based upsampling strategy and a controllable speech synthesis paradigm based on pitch and energy prediction.
Subsequently, many recent models, such as the StyleTTS~\cite{StyleTTS2} and NaturalSpeech series~\cite{naturalspeech3} achieve more natural speech synthesisby incorporating techniques such as diffusion models and adversarial training.
Meanwhile, speech synthesis models based on discrete speech codecs and autoregressive architectures have also emerged progressively, such as SparkTTS~\cite{sparktts}, Llasa~\cite{llasa}, and CosyVoice series~\cite{cosyvoice3}.
Despite the progress, they cannot be directly applied to movie dubbing tasks because they lack the design of modeling duration and prosody from performance in the given movie clips.

\subsection{Visual Voice Cloning} 
Some previous V2C methods attempt to improve dubbing quality by pretraining the phoneme encoder~\cite{spk2dub} or decoupling acoustic modeling and prosody adaptation~\cite{ProDubber}, in response to the scarcity and noisiness of movie dubbing datasets caused by issues such as copyright constraints.
Another group of work explores techniques such as flow matching~\cite{emodubber} and contrastive learning~\cite{styledubber}, primarily aiming to enhance the performance of audiovisual alignment~\cite{HPMDubbing, zhao2025towards, cong2025flowdubber, li2025dubbing}.
However, their visual-based alignment methods struggle to generalize to unseen video scenarios, hindering the broader application.


\subsection{Dubbing with MLLMs} 
Multimodal large language models (MLLMs) have powerful and generalizable capabilities for multimodal content understanding.
The earliest works leveraging MLLMs for dubbing primarily utilized them to align the overall duration between the dubbing and the video for multilingual dubbing~\cite{dubwise}.
Recently, VoiceCraft-Dub~\cite{voicecraftdub} attempted to autoregressively generate dubbing by feeding video frames and script text into an LLM-decoder. DeepDubber~\cite{deepdubber} attempts to leverage a chain-of-thought prompting strategy to guide the MLLMs generating coarse-grained information—such as scene type, character age, and gender—from the video to assist alignment.
However, they overlook the capability of fine-grained MLLM-generated dubbing instructions in alignment and their potential in zero-shot dubbing.

%% file: Sections/3-Method.tex
\section{Method}

\subsection{Overview}
The target of the overall movie dubbing task is:
\begin{equation}
     \hat{A} _{Dub} = \mathrm{Model}(A_{Ref}, T_{d}, V_{m}),
\end{equation}
where the $\hat{A} _{Dub}$ is the generated dubbing and $A_{Ref}, T_{d}, V_{m}$ are the reference audio, dubbing scripts, and the input silent movie clip, respectively.
The core of audio-visual alignment in movie dubbing lies in assigning accurate phoneme-level durations and prosody attributes based on the visual content and the corresponding dubbing script:
\begin{equation}
     \tilde{p}, \tilde{n}, \tilde{d} = \mathrm{Alignment}(T_d,V_{m}),
\end{equation}
where the $\tilde{p}$, $\tilde{n}$, and $\tilde{d}$ are predicted pitch, energy, and duration that align with the video content.

Figure~\ref{method} illustrates the framework of our proposed instruction-based alignment approach.
We employ a pre-trained multimodal large language model to generate natural language fine-grained instructions of the character's speaking rate and emotional dynamics by taking both the movie clip and the script as input with the corresponding prompt.
Then, the instructed duration distilling module mines the duration cues from the speaking rate instructions, which are then combined with prosodic text features to predict phoneme-level durations $\tilde{d}$.
Meanwhile, the instructed emotion calibrating module fine-tunes a lightweight LLM-based analyzer to extract the dubbing emotions from emotion instructions, thereby guiding the prediction of emotion-aligned prosody $\tilde{p}$ and $\tilde{n}$.
We detail each module below.

\subsection{Instructed Duration Distilling}
To enable the MLLM to generate fine-grained instructions that capture the character’s speaking rate in the input video, we feed both the video clip, movie script, and the speaking rate prompt into the MLLM as input:
\begin{equation}
     I_{Dur} = \mathrm{MLLM}(T_d,V_{m},Prompt_{Dur}),
\end{equation}
where $I_{Dur}$ is the fine-grained speaking rate instruction of the given video, $Prompt_{Dur}$ is the speaking rate prompt.
Compared to conventional methods that extract visual features from each video frame, the fine-grained instructions generated by MLLMs are more informative, yet often contain redundant elements such as prepositions and conjunctions.
Therefore, it is essential to distill the discriminative duration cues from them for accurate duration prediction.

To alleviate this problem, we propose the Instructed Duration Distilling (IDD) module, which consists of multiple distilling layers that leverage the slot attention mechanism~\cite{slotattention}.
Slot attention initializes a set of learnable prototypes, referred to as element slots, which interact with sequence inputs to iteratively group information corresponding to the same underlying element. 
Within the IDD module, it is particularly well-suited for extracting discriminative duration cues from fine-grained speaking rate instructions, enabling effective alignment modeling.

Specifically, we first employ a global text embedding (GTE) module to convert the speaking rate instruction $I_{Dur}$ into text embeddings $E_{Dur}$ with position encoding:
\begin{equation}
     E_{Dur} = \mathrm{GTE}(T_{Dur})\in \mathbb{R}^{L_{Dur}\times d_{GTE}}.
\end{equation}
The input features of the distilling layers $E_{Dur}$ are first linearly projected to obtain key and value representations:
\begin{equation}
     K_{Dur} = W^KE_{Dur}, V_{Dur} = W^VE_{Dur}, 
\end{equation}
where $W^K$ and $W^V \in \mathbb{R}^{d_{GTE}} $ are learnable linear projections.
A fixed number $K$ of duration prototype slots $\{s^{(0)}_k\}^K_{k=1}$ are initialized randomly as learned duration features shared across all inputs.
For $T$ iterations (\ie, $T$ distilling layers), each slot $s_k^{(t)}$ is updated by attending to the input features. 
At each iteration $t$, the current slots are projected into queries:
\begin{equation}
     Q^{(t)}=W^QS^{(t)}, 
\end{equation}
where $S^{(t)} = [s_1^{(t)};...;s_K^{(t)}]$ and $W^Q$ is a learnable projection.
Attention weights between each slot and input token are computed using scaled dot-product attention:
\begin{equation}
     \alpha_{k,n}^{(t)}=\frac{\exp\left(\frac{q_k^{(t)}\cdot k_n}{\sqrt{d}}\right)}{\sum_{k^{\prime}=1}^K\exp\left(\frac{q_k^{(t)}\cdot k_n}{\sqrt{d}}\right)},\quad\mathrm{for~}k=1\ldots K,
\end{equation}
where $q_k^{(t)}\in \mathbb{R}^{d_{GTE}}$ is the $k$-th query vector and $k_n \in \mathbb{R}^{d_{GTE}}, n=1,\ldots,L_{Dur}$ is the $n$-th key vector.
Then, each slot $s_k^{(t)}$ receives the weight summary of the input values and updated via a GRU-based mechanism:
\begin{equation}
\begin{split}
    u_k^{(t)}&=\sum_{n=1}^N\alpha_{k,n}^{(t)}\cdot v_n,\quad v_n\in\mathbb{R}^{d_{GTE}},
    \\ s_k^{(t+1)}&=\mathrm{GRU}(u_k^{(t)},s_k^{(t)})+\mathrm{MLP}(\mathrm{LN}(s_k^{(t+1)})),
    \end{split}
\end{equation}
where $\mathrm{LN}$ denotes layer normalization, and $\mathrm{MLP}$ is a small feedforward network with non-linearity.
After $T$ iterations ($T$ distilling layers with shared weights), we obtain the final slots $S^{(T)}$ as the duration features $P_{Dur}$, which are distilled from the speaking rate instructions:
\begin{equation}
     P_{Dur} = S^{(T)} \in \mathbb{R}^{K\times d_{GTE}}.
\end{equation}

After getting the distilled duration features, following~\cite{ProDubber}, we convert the script text into phonemes and extract prosodic text features $T_p$ using a pretrained phoneme-level BERT model~\cite{pl=bert}:
\begin{equation}
\begin{split}
    T_{pho} &= \mathrm{G2P}(T_d)\in \mathbb{R}^{L_{pho}},
    \\ T_p &= \mathrm{BERT}_{pho}(T_{pho})\in \mathbb{R}^{L_{pho}\times d_{m}},
    \end{split}
\end{equation}
where the $\mathrm{G2P}$ and $\mathrm{BERT}_{pho}$ are the grapheme to phoneme transfer and phoneme-level BERT. 
We use the $T_p$  as queries, while the dimension-reduced duration features $P^{\prime}_{Dur}$ as both keys and values to a cross-attention (CA) layer. 
The fused duration representations are then fed into a duration predictor to obtain the final duration output:
\begin{equation}
\begin{split}
    F_{Dur}&=\mathrm{CA}(T_p, P_{Dur}^{\prime}, P_{Dur}^{\prime})\in\mathbb{R}^{L_{pho}\times d_{m}},
    \\ \tilde{d} &= \mathrm{DurationPrdictor}(F_{Dur})\in\mathbb{R}^{L_{pho}},
    \end{split}
\end{equation}
where the $\mathrm{DurationPrdictor}$ is a Bi-LSTM network with a prediction head following~\cite{StyleTTS2}.
The $ \tilde{d}$ is scaled according to the length of input video to ensure the consistency between total duration of video and dubbing.

\subsection{Instructed Emotion Calibrating}
First, we input the video, script, and prompt into the MLLM to obtain fine-grained emotion instructions:
\begin{equation}
     I_{Emo} = \mathrm{MLLM}(T_d,V_{m},Prompt_{Emo}).
\end{equation}
Similarly, it is essential to extract discriminative emotional variations from fine-grained emotion instructions to facilitate accurate emotion-aligned prosody prediction of each phoneme.
To this end and also to enhance the model's generalization ability in emotion analysis, we introduce a predefined set of emotions and use the elements as emotion entities.
After a comprehensive analysis of several emotion-labeled speech and dubbing datasets~\cite{V2C, esd}, we adopt the following seven emotions as the predefined set of emotion entities: happy, angry, disgust, fear, neutral, sad, and surprise.

We employ a lightweight LLM as the emotion instruction analyzer to extract the emotion entities from the emotion instructions.
Introducing an additional analyzer during training helps prevent the MLLM's knowledge from being biased by the visual styles of specific dubbing datasets, thereby preserving the model's generalization capability while also enhancing its flexibility to accommodate different MLLMs.
To ensure consistency between the emotion entities extracted from the emotion instruction and those present in the ground truth dubbing, we employ an audio large language model (Audio LLM) to extract ground truth emotion entities from the reference dubbing:
\begin{equation}
     Entity_{Emo} = \mathrm{AudioLLM}(A_{Dub},Prompt_{Entity}),
\end{equation}
where $Entity_{Emo}$ denotes the ground-truth emotion entities extracted from the dubbing audio.

\input{Tables/Ori}

Supervised by ground truth emotion entities, we fine-tune this analyzer using Low Rank Adaptation (LoRA)~\cite{lora} to calibrate its analysis of the emotion instruction.
For each QKV projection layer and feed-forward layer in the emotion instruction analyzer, we perform fine-tuning using rank-$R$ adaptation matrices:
\begin{equation}
    W^{\prime}=W+\Delta W=W+AB,
\end{equation}
where $W$ is the original parameter, $A\in\mathbb{R}^{d_{LLM}\times R}$ and $B\in\mathbb{R}^{R\times d_{LLM}}$ together constitute the complete set of trainable parameters $\theta$. 
During training, the parameters $\theta$ are optimized by minimizing the autoregressive loss:
\begin{equation}
    \theta^{\prime}\leftarrow\theta-\eta\cdot\nabla_\theta\mathcal{L}(\mathrm{Analyzer}(I_\mathrm{Emo};\theta),Entity_{Emo}),
\end{equation}
where $\eta$ is the learning rate.

After the fine-tuning, we use the emotion instruction analyzer to extract emotion entities from the emotion instructions and obtain their corresponding text embeddings:
\begin{equation}
     E_{Emo} = \mathrm{GTE}(\mathrm{Analyzer}(I_{Emo},Prompt_{A}, \theta^{\prime})),
\end{equation}
where $Prompt_{A}$ is the analysis prompt, $E_{Emo}\in\mathbb{R}^{L\times d_{GTE}}$ is the text embedding of the predicted emotion entities.
Then, we apply cross-attention between the dimension-reduced emotion features $E_{Emo}^{\prime}$ and the prosodic text features to obtain prosody features for prosody prediction:
\begin{equation}
\begin{split}
    F_{Emo}&=\mathrm{CA}(T_p, E_{Emo}^{\prime}, E_{Emo}^{\prime})\in\mathbb{R}^{L_{pho}\times d_{m}},
    \\ \tilde{p}, \tilde{n}&= \mathrm{ProsodyPrdictor}(F_{Emo})\in\mathbb{R}^{L_{pho}},
    \end{split}
\end{equation}
where the $\mathrm{ProsodyPrdictor}$ has the same architecture as the duration predictor with two prediction heads for pitch and energy, respectively.

\subsection{Audio Generation and Training Objective}
We feed the predicted duration and prosody, along with the script text and reference audio, into a pre-trained HiFi-GAN-based audio decoder~\cite{hifigan} to generate the final dubbing audio:
\begin{equation}
     \hat{A} _{Dub} = \mathrm{AudioDecoder}(T_{d}, \tilde{p}, \tilde{n}, \tilde{d}, A_{Ref}).
\end{equation}
The overall training objective of InstructDubber is:
\begin{equation}
    \mathcal{L}_{total} = \lambda _1\mathcal{L}_{Dur}+\lambda _2\mathcal{L}_{Emo},
    \label{loss}
\end{equation}
\begin{equation}
    \mathcal{L}_{Dur} = \frac{1}{L_{pho}} \sum_{i=0}^{L_{pho}-1} \left \| \widetilde d_i-d_i \right \|_1,
\end{equation}
\begin{equation}
\mathcal{L}_{Emo} = \frac{1}{L_{pho}} \sum_{i=0}^{L_{pho}-1} \left \| \widetilde p_i-p_i \right \|_1 + \left \| \widetilde n_i-n_i \right \|_1,
\end{equation}
where the weight in Equation~\eqref{loss} are set to $\lambda_1 = 2$, $\lambda_2 = 1$.
We use the ground truth emotion entities during training and the predicted during inference and evaluation.

%% file: Tables/Ori.tex
\begin{table*}[!t]
  \centering

  \resizebox{1.0\linewidth}{!}
  {

    \begin{tabular}{c|cccc|cccc|cccc|}
    \hline
    
    Benchmark & \multicolumn{4}{c|}{V2C-Animation} & \multicolumn{4}{c|}{Chem} & \multicolumn{3}{c}{GRID} \\
    \midrule
    Methods 
    & DD $\downarrow$ 
    & \small{EMO-SIM} (\%) $\uparrow$
    & WER (\%) $\downarrow$
    & \small{UTMOS} $\uparrow$   
    & DD $\downarrow$ 
    & \small{EMO-SIM} (\%) $\uparrow$
    & WER (\%) $\downarrow$
    & \small{UTMOS} $\uparrow$   
    & DD $\downarrow$ 
    & WER (\%) $\downarrow$
    & \small{UTMOS} $\uparrow$     \\
    \midrule
    GT  & 0.0000 & 100.00 & 25.55 & 2.26 & 0.0000 & 100.00 & 3.85 & 4.18 & 0.0000 & 22.41 & 3.94\\
    \midrule
    Speak2Dub~\cite{spk2dub} & 0.5173 & 66.58 & 17.51 & 2.41 & 0.4786 & 76.78 & 11.82 & 3.72 & 0.2650 & \textbf{17.40} & 3.69 \\
    StyleDubber~\cite{styledubber} & \textbf{0.5092} & \underline{67.22} & 31.94 & 1.89 & \underline{0.4508} & \underline{77.99} & 13.14 & 3.02 & \textbf{0.2453} & 18.88 & 3.73 \\
    DeepDubber*~\cite{deepdubber} & 0.5756 & 56.42 & 35.88 & 2.03 & 0.5041 & 52.37 & 25.51 & 2.53 & 0.3995 & 51.16 & 2.31 \\
    ProDubber~\cite{ProDubber} & 0.5148 & 67.15 & \textbf{8.04} & \underline{3.10} & 0.4673 & 76.69 & \underline{9.45} & \underline{3.85} & 0.2551 & 18.60 & \underline{3.87} \\
    \midrule
    InstructDubber (Ours) & \underline{0.5122} & \textbf{68.46} & \underline{9.27} & \textbf{3.11} & \textbf{0.4461} & \textbf{78.38} & \textbf{8.86} & \textbf{3.87} & \underline{0.2522} & \underline{17.81} & \textbf{3.88} \\
    \bottomrule
    \end{tabular}
    }
        \caption{
  Results of in-domain dubbing on three major benchmarks, which use the train set and test set from the same benchmark for training and evaluation. The best results are \textbf{in bold} and the second-best ones are \underline{underlined}.
  }

  \label{Ori_result}%
\end{table*}%

%% file: Sections/4-Experiment.tex
\section{Experiments}

\subsection{Datasets}
\textbf{V2C-Animation dataset~\cite{V2C}} is a collection of 10,217 video clips from 26 animated movies,
consists of 10,217 video-audio-text triplets cropped from 26 Disney animated movies, totaling 153 different speakers, with complete speaker and emotion annotations.

\noindent\textbf{Chem dataset~\cite{chem_dataset}} is a popular dubbing dataset recording a chemistry teacher speaking in the class. 
For complete dubbing, each video has clip to sentence-level following~\cite{nerualdubber}. %

\noindent \textbf{GRID dataset~\cite{GRID}} is a multi-speaker dubbing benchmark which
comprises video recordings of 33 speakers performing 1,000 scripted sentences each.


\subsection{Evaluation Metrics}
\noindent \textbf{Duration Divergence (DD).}
The duration divergence evaluates the lip-synchronization by calculating the divergence between phoneme-level duration distributions of synthesized and ground truth dubbing following~\cite{flashspeech}.

\noindent \textbf{EMO-SIM.}
Emotion similarity (EMO-SIM) measures the cosine similarity between the emotion embedding of generated dubbing and ground truth, which is obtained using Emotion2Vec~\cite{emotion2vec} following~\cite{voicecraftdub}.
Since GRID videos predominantly feature neutral expressions, we only conduct EMO-SIM on V2C-Animation and Chem benchmarks.

\noindent \textbf{WER.}
The Word Error Rate (WER)~\footnote{https://github.com/jitsi/jiwer} assesses the model's pronunciation accuracy by using an advanced ASR model Whisper\footnote{https://huggingface.co/openai/whisper-large}~\cite{whisper} to transcribe the dubbing into text and compare it with the original dubbing script.

\noindent \textbf{UTMOS.}
UTMOS~\cite{utmos} is a speech mean opinion score (MOS) predictor to measure the acoustic quality and naturalness of the generated dubbing following~\cite{naturalspeech3}.

\input{Tables/Cross}
\input{Tables/Cross_2}


\subsection{Implementation Details}
We employ a pre-trained JDC network~\cite{JDC} as the pitch extractor and use the log norm to calculate the energy following~\cite{StyleTTS2}.
To get the ground truth duration and calculate the duration divergence metrics, we 
adopt the ASR model fine-tuned for the TTS task as text aligner~\cite{styletts} to get the alignment between phoneme and mel-spectrogram following StyleTTS2~\cite{StyleTTS2}.
For audio generation, we adopt the same pre-trained audio decoder as ProDubber~\cite{ProDubber}.
We use the Qwen2.5-Instruct-7B~\cite{qwen2.5} to analyze the emotion instructions and the Qwen2.5-Omni-7B~\cite{Qwen2.5-Omni} to get ground truth emotion entities.
An Adam~\cite{adam} with $\beta_1=0.9$, $\beta_2=0.98$, $\epsilon=10^{-9}$ is used as the optimizer during the training.
The learning rate is set to 0.00625.

\subsection{Comparison with SOTA Methods}
\subsubsection{Results on in-domain dubbing.}
As shown in Table~\ref{Ori_result}, InstructDubber outperforms existing state-of-the-art models on the majority of evaluation metrics across three dubbing benchmarks.
In terms of lip synchronization, InstructDubber achieves the lowest duration divergence on Chem benchmarks, and second best on V2C-Animation and GRID benchmarks.
It indicates that InstructDubber can generate dubbing exhibits the smallest temporal discrepancy with the ground truth.
The highest EMO-SIM on both the V2C-Animation and Chem benchmarks shows that the dubbing generated by InstructDubber exhibits the closest emotional expressiveness to the ground truth, demonstrating the best emotion alignment performance in this scenario.

Moreover, due to natural language instructions being inherently less susceptible to noise compared to visual features, resulting in more stable predictions of duration and prosody, the pronunciation clarity and dubbing quality of the generated dubbing are also improved.
We achieve competitive WER performance across all three benchmarks and observe an improvement in UTMOS, indicating the best overall dubbing alignment and quality.

\subsubsection{Results on zero-shot dubbing.}
We conduct pairwise zero-shot dubbing evaluations across the three benchmarks by directly using unseen videos from another dataset for evaluation.
As shown in Table~\ref{cross_1} and~\ref{cross_2}, InstructDubber outperforms state-of-the-art models across all six zero-shot dubbing scenarios on both alignment accuracy and dubbing quality.

Specifically, compared to previous approaches that rely on visual features to achieve alignment, which are easily perturbed by variations in visual domains, guidance derived from natural language instructions remains invariant to visual domain variations, enabling more accurate and robust alignment in temporal and emotional aspects.
More accurate prediction of duration and prosody also leads to significant improvements in pronunciation clarity and speech quality of the generated dubbing.
Besides, compared to the DeepDubber, which uses coarse-grained instructions as a supplement to visual features and is trained across multiple benchmarks, our proposed instructed duration distilling and instructed emotion calibrating module effectively leverages fine-grained dubbing instructions to guide the prediction of aligned duration and prosody, consistently achieving superior performance in all six zero-shot dubbing scenarios.

\input{Tables/ablation}

\begin{figure*}[t]
 \centering
  \resizebox{\linewidth}{!}{\includegraphics{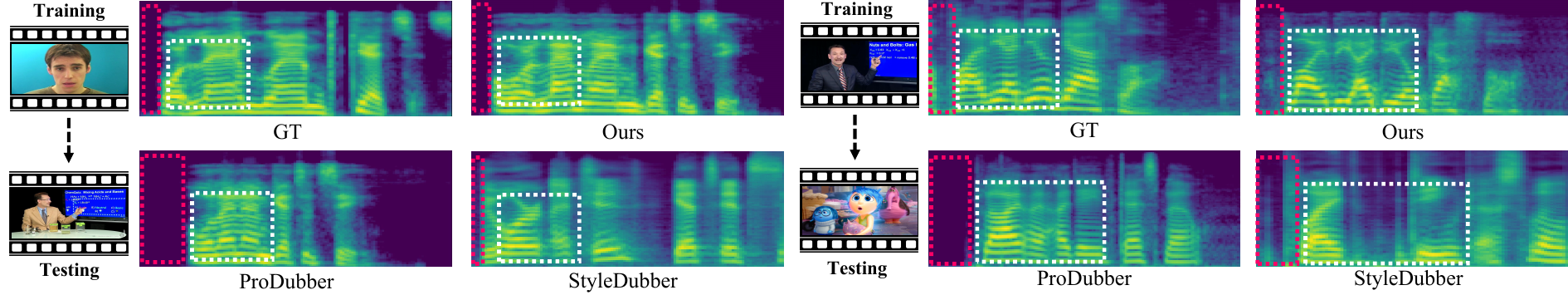}}
  \caption{
  The mel-spectrograms of ground truth and synthesized dubbing by different models in zero-shot scenario.
  The red and white boxes highlight regions where different models exhibit significant differences in temporal and prosody alignment.
  }
  \label{visual}
\end{figure*}

\subsection{Ablation Studies}
To validate the effectiveness of each proposed module, we conduct ablation studies on V2C-Animation, Chem, and their cross-domain zero-shot scenarios (\ie, V2C2Chem and Chem2V2C)
.
We report the average performance of the four scenarios in this section.

\subsubsection{Ablation of each module.}
We integrate the proposed Instructed Duration Distilling (IDD) and Instructed Emotion Calibrating (IEC) module into the visual feature-based dubbing baseline model separately to validate their effectiveness.
As shown in Table~\ref{abl}, incorporating the IDD module leads to improved duration alignment performance, with the average duration divergence (DD) reduced compared to the baseline model.
More accurate duration alignment also leads to clearer pronunciation and improved dubbing quality, resulting in performance gains in both WER and UTMOS.
The IEC module enables the model to better predict emotion-aligned prosody, achieving +1.43\% performance gain on the EMO-SIM metric.
The two proposed modules together enable InstructDubber to achieve the best overall dubbing performance, validating the effectiveness of both components.

\input{Tables/ablation_duration}
\input{Tables/ablation_emotion}

\subsubsection{Ablation of instruction-based duration alignment.}
In Table~\ref{abl_duration}, we report the performance of various strategies to leverage the speaking rate instruction for duration alignment.
Directly applying cross-attention to fine-grained speaking rate instruction and dubbing script (Raw Instruction CA) fails to extract discriminative duration cues from the instruction, resulting in degraded duration alignment performance.
Finetuning an LLM using LoRA to directly predict the duration sequence based on the instruction (LoRA Prediction) suffers from implicit duration value optimization and unstable inference success rate.
Instead, the proposed IDD module achieves the best lip-synchronization performance by mining duration cues from fine-grained instructions.
Through an ablation on the number of learnable duration prototypes, we find that the 10-prototype achieves better balance on alignment accuracy and speech quality.

\subsubsection{Ablation of instruction-based emotion alignment.}
The results of ablation on instruction-based emotion alignment are shown in Table~\ref{abl_emotion}.
Directly applying fine-grained emotion instructions in a cross-attention mechanism to predict prosody introduces detailed yet redundant information, which leads to suboptimal performance on the EMO-SIM metric and dubbing quality.
Using the instruction distilling method similar to the IDD module or directly extracting emotion entities from the instructions lacks supervision of emotion state from the ground truth dubbing audio, limiting the accuracy in reflecting the character’s emotional state in the video.
The proposed IEC module incorporates ground truth emotion entities as supervision to calibrate the analysis of emotion instruction, which improves the emotion alignment and achieves superior performance on EMO-SIM.

\input{Tables/ablation_GTE}

\subsubsection{Ablation of different MLLMs and GTE models.}
To validate the robustness of our proposed method
, we conducted ablation experiments using different MLLMs (VideoLLaMA3-7B~\cite{videollama3} and LLaVA-NEXT-7B~\cite{llavanext}) and GTE models of varying sizes~\cite{qwen3-embedding}.
Table~\ref{abl_gte} shows that InstructDubber achieves advanced and stable performance
under different configurations of MLLM and GTE size.

\subsection{Qualitative Analysis}
We visualize ground-truth and zero-shot dubbing mel-spectrograms from different models in Figure~\ref{visual}. 
As shown in the red box, our model produces more accurate phoneme-level duration predictions, leading to superior lip synchronization. 
The white box further highlights spectrogram patterns and variations which shows that we achieves prosody patterns more consistent with the ground truth.

%% file: Tables/Cross.tex
\begin{table*}[!t]
  \centering

  \resizebox{1.0\linewidth}{!}
  {

    \begin{tabular}{c|cccc|ccc|cccc}
    \hline
    
    Benchmark & \multicolumn{4}{c|}{V2C2Chem} & \multicolumn{3}{c|}{V2C2GRID} & \multicolumn{4}{c}{GRID2V2C} \\
    \midrule
    Methods 
    & DD $\downarrow$ 
    & \small{EMO-SIM} (\%) $\uparrow$
    & WER (\%) $\downarrow$
    & \small{UTMOS} $\uparrow$   
    & DD $\downarrow$ 
    & WER (\%) $\downarrow$
    & \small{UTMOS} $\uparrow$   
    & DD $\downarrow$ 
    & \small{EMO-SIM} (\%) $\uparrow$
    & WER (\%) $\downarrow$
    & \small{UTMOS} $\uparrow$     \\
    \midrule
    GT  & 0.0000 & 100.00 & 3.85 & 4.18 & 0.0000 & 22.41 & 3.94 & 0.0000 & 100.00 & 25.55 & 2.26\\
    \midrule
    Speak2Dub & 0.5334 & 57.21 & \underline{10.23} & 2.66 & 0.3258 & 64.31 & 2.91 & 0.5506 & 36.96 & 65.82 & 2.52 \\
    StyleDubber & 0.4721 & 67.03 & 18.63 & 2.26 & 0.3380 & 77.97 & 2.58 & 0.5397 & 39.92 & 70.32 & 1.93 \\
    DeepDubber* & 0.5041 & 52.37 & 25.51 & 2.53 & 0.3995 & 51.16 & 2.31 & 0.5756 & \underline{56.42} & \underline{35.88} & 2.03 \\
    ProDubber & \underline{0.4649} & \underline{68.21} & 10.32 & \underline{3.62} & \underline{0.3146} & \underline{25.67} & \underline{3.55} & \underline{0.5367} & 51.60 & 41.27 & \underline{2.73} \\
    \midrule
    Ours & \textbf{0.4565} & \textbf{70.34} & \textbf{8.42} & \textbf{3.71} & \textbf{0.3103} & \textbf{23.49} & \textbf{3.61} & \textbf{0.5261} & \textbf{57.09} & \textbf{19.56} & \textbf{2.85} \\
    \bottomrule
    \end{tabular}
    }
\caption{
  Results on zero-shot movie dubbing across three major benchmarks. 
  For example, V2C2GRID indicates that using the checkpoint trained on the V2C-Animation dataset to directly dub the video clip from GRID dataset without any fine-tuning.
  Note that the official checkpoint of DeepDubber* is trained jointly on multiple datasets, including V2C-Animation and GRID, and therefore exhibits identical performance across various zero-shot settings and as reported in Table~\ref{Ori_result}.
  }
  \label{cross_1}%
\end{table*}%

%% file: Tables/Cross_2.tex
\begin{table*}[!t]
  \centering
  \resizebox{1.0\linewidth}{!}
  {

    \begin{tabular}{c|cccc|ccc|cccc}
    \hline
    
    Benchmark & \multicolumn{4}{c|}{Chem2V2C} & \multicolumn{3}{c|}{Chem2GRID} & \multicolumn{4}{c}{GRID2Chem} \\
    \midrule
    Methods 
    & DD $\downarrow$ 
    & \small{EMO-SIM} (\%) $\uparrow$
    & WER (\%) $\downarrow$
    & \small{UTMOS} $\uparrow$   
    & DD $\downarrow$ 
    & WER (\%) $\downarrow$
    & \small{UTMOS} $\uparrow$   
    & DD $\downarrow$ 
    & \small{EMO-SIM} (\%) $\uparrow$
    & WER (\%) $\downarrow$
    & \small{UTMOS} $\uparrow$     \\
    \midrule
    GT  & 0.0000 & 100.00 & 25.55 & 2.26  & 0.0000 & 22.41 & 3.94 & 0.0000 & 100.00 & 3.85 & 4.18 \\
    \midrule
    Speak2Dub & 0.5873 & 59.72 & 23.78 & 2.74 & \underline{0.3123} & 55.08 & 3.00 & 0.5832 & 35.14 & 60.47 & 2.58 \\
    StyleDubber & \underline{0.5627} & 58.54 & 25.43 & 1.95 & 0.3139 & 67.46 & 2.10 & 0.5095 & 41.35 & 68.91 & 2.05 \\
    DeepDubber* & 0.5756 & 56.42 & 35.88 & 2.03 & 0.3995 & 51.16 & 2.31 & \underline{0.5041} & 52.37 & \underline{25.51} & 2.53 \\
    ProDubber & 0.5650 & \underline{65.98} & \underline{14.33} & \underline{2.91} & 0.3209 & \underline{47.42} & \underline{3.73} & 0.5781 & \underline{54.87} & 30.17 & \underline{2.72} \\
    \midrule
    Ours & \textbf{0.5583} & \textbf{66.57} & \textbf{12.60} & \textbf{3.07} & \textbf{0.3042} & \textbf{38.53} & \textbf{3.84} & \textbf{0.4849} & \textbf{58.91} & \textbf{20.73} & \textbf{2.94} \\
    \bottomrule
    \end{tabular}
    }
    \caption{
  Results on zero-shot movie dubbing across three major benchmarks with same zero-shot setting as Table~\ref{cross_1}.
  }
  
  \label{cross_2}%
\end{table*}%

%% file: Tables/ablation.tex
\begin{table}[!t]
\centering
\resizebox{1.0\linewidth}{!}{
\begin{tabular}{c|cccc}
    \hline
    Methods & DD $\downarrow$ & \small{EMO-SIM} (\%) $\uparrow$ & WER(\%) $\downarrow$ & UTMOS $\uparrow$\\
    \toprule
    Visual Feature & 0.5030 & 69.51 & 10.54 & 3.37 \\
    \midrule
    w/ IDD  & \underline{0.4941} & 67.85 & \underline{9.97} & \underline{3.42}\\
    w/ IEC  & 0.5046 & \underline{70.77} & 11.76 & 3.40 \\
    \midrule
    Full Model & \textbf{0.4933} & \textbf{70.94} & \textbf{9.79} & \textbf{3.44} \\
    \bottomrule
\end{tabular}}
\caption{Results of ablation study on each module.}
\label{abl}
\end{table}

%% file: Tables/ablation_duration.tex

\begin{table}[!t]
\centering
\resizebox{1.0\linewidth}{!}{
\begin{tabular}{c|cccc}
    \hline
    Method & DD $\downarrow$  & WER(\%) $\downarrow$ & UTMOS $\uparrow$\\
    \toprule
    Raw Instruction CA & 0.5072 & 11.55 & 3.40\\
    LoRA Prediction & 0.5331 & 13.58 & 3.32 \\
    \midrule
    IDD (Prototype Num = 1)  & 0.5005 & 11.63 & 3.38 \\
    IDD (Prototype Num = 5)  & \underline{0.4975} & \textbf{9.23} & 3.43\\
    IDD (Prototype Num = 10)  & \textbf{0.4933} & \underline{9.79} & \underline{3.44} \\
    IDD (Prototype Num = 20)  & 0.4991  & 10.16 & \textbf{3.45}\\
    \bottomrule
\end{tabular}}
\caption{Results of ablation study on different instruction-based duration alignment methods.}
\label{abl_duration}
\end{table}

%% file: Tables/ablation_emotion.tex
\begin{table}[!t]
\centering
\resizebox{1.0\linewidth}{!}{
\begin{tabular}{c|cccc}
    \hline
    Method   &  \small{EMO-SIM} (\%) $\uparrow$  & WER(\%) $\downarrow$ & UTMOS $\uparrow$\\
    \toprule
    Raw Instruction CA & \underline{69.55} & 10.41 & 3.38 \\
    Instrcution Distilling & 70.51 & \underline{9.92} & \underline{3.42}\\
    w/o Calibrating & 70.33 & 10.52 & 3.39\\
    \midrule
    Ours (IEC) & \textbf{70.94} & \textbf{9.79} & \textbf{3.44}\\
    \bottomrule
\end{tabular}
}
\caption{Results of ablation study on different instruction-based emotion alignment methods.}

\label{abl_emotion}
\end{table}

%% file: Tables/ablation_GTE.tex
\begin{table}[!t]
\centering
\resizebox{1.0\linewidth}{!}{
\begin{tabular}{c|cccc}
    \hline
    Models & DD $\downarrow$ & \small{EMO-SIM} (\%) $\uparrow$ & WER(\%) $\downarrow$ & UTMOS $\uparrow$\\
    \toprule
    VideoLLaMA3-7B & 0.4924 & 70.09 & 9.84 & 3.46\\
    LLaVA-Next-7B &  0.4933 & 70.94 & 9.79 & 3.44\\
    \midrule
    GTE-0.6B& 0.4988 & 70.99 & 10.02 & 3.43 \\
    GTE-1.5B  & 0.4933 & 70.94 & 9.79 & 3.44 \\
    GTE-4B & 0.4999 & 70.51 & 10.27 & 3.41 \\
    \bottomrule
\end{tabular}}
\caption{Results of ablation study on different MLLMs and GTE models of varying size.}
\label{abl_gte}
\end{table}

%% file: Sections/5-Conclusion.tex
\section{Conclusion}
In this paper, we propose InstructDubber, a novel instruction-based alignment dubbing method that achieves robust both in-domain and zero-shot dubbing.
The proposed instructed duration distilling module and the instructed emotion calibrating effectively leverage fine-grained dubbing instructions to facilitate temporal and emotional alignment, respectively.
The superior performance on both in-domain and zero-shot experiments across three major benchmarks, along with the ablation studies, demonstrates the effectiveness of the proposed method and each module.

\section*{Acknowledgements}
 \begin{sloppypar}
This work was supported by the National Nature Science Foundation of China (62322211, U21B2024), the ``Pioneer'' and ``Leading Goose'' R\&D Program of Zhejiang Province(2024C01023, 2024C01107, 2023C01030, 2023C03012), Key Laboratory of Intelligent Processing Technology for Digital Music (Zhejiang Conservatory of Music), Ministry of Culture and Tourism (2023DMKLB004).
Yuankai Qi and Tao Gu are not supported by the aforementioned fundings.
\end{sloppypar}